\documentclass{vienna-conf2019}
\usepackage{graphicx}
\usepackage{hyperref}
\usepackage[]{natbib}  
\usepackage{epstopdf}

\def\BibTeX{{\rm B\kern-.05em{\sc i\kern-.025em b}\kern-.08em
    T\kern-.1667em\lower.7ex\hbox{E}\kern-.125emX}}
\bibpunct{(}{)}{;}{a}{}{,}  

\begin{document}

\TitreGlobal{Stars and their variability observed from space}


\title{From the Sun to solar-like stars: how does the solar
modelling problem affect our studies of solar-like
oscillators?}

\runningtitle{From the Sun to solar-like stars}

\author{G. Buldgen}\address{Observatoire de Gen\`eve, Universit\'e de Gen\`eve, 51 Ch. Des Maillettes, CH−1290 Sauverny, Suisse}

\author{C. Pezzotti$^1$}

\author{M. Farnir}\address{STAR Institute, Universit\'e de Li\`ege, All\'ee du Six Ao\^ut 19C, B−4000 Li\`ege, Belgium}

\author{S.J.A.J. Salmon$^2$}

\author{P. Eggenberger$^1$}

\setcounter{page}{237}


\maketitle


\begin{abstract}
Since the first observations of solar oscillations in 1962, helioseismology has probably been one of the most successful fields of astrophysics. Besides the improvement of observational data, solar seismologists developed
sophisticated techniques to infer the internal structure of the Sun.
Back in 1990s these comparisons showed a very high agreement between solar models and the Sun. However, the downward revision of the CNO surface abundances in the Sun in 2005, confirmed in 2009, induced a drastic reduction of this agreement leading to the so-called solar modelling problem. More than ten years later, in the era of the space-based photometry missions which have established asteroseismology of solar-like stars as a standard approach to obtain their masses, radii and ages, the solar modelling problem still awaits a solution. I will briefly present the results of new helioseismic inversions, discuss the current uncertainties of
solar models and possible solutions to the solar modelling problem. I will also discuss how the solar problem can have significant implications for asteroseismology as a whole by discussing the modelling of the exoplanet-host star Kepler-444, thus impacting the fields requiring a precise and accurate knowledge of stellar masses, radii and ages, such as Galactic archaeology and exoplanetology.
\end{abstract}



\section{Introduction}

The advent of space based photometry missions such as CoRoT \citep{Baglin}, \textit{Kepler} \citep{Borucki}, TESS \citep{Ricker2015} and the BRITE constellation \citep{Weiss2014} has enabled us to thoroughly test the physical ingredients of theoretical stellar models using asteroseismology. In this era of precision stellar physics, various modelling strategies can be adopted to derive fundamental stellar parameters, e.g. fitting of individual frequencies and frequency combinations or of global asteroseismic indices. Ultimately, the derived precision and accuracy of such inferences is not limited by the propagation of the observational uncertainties on the stellar properties, but by the limitations of theoretical stellar models. 

A good illustration of the current shortcomings of stellar models is the so-called ``solar modelling problem'' which stems from the downward revision of the solar metallicity by \cite{AGSS09}. In this context, we show that varying the ingredients entering the standard solar model while still following its framework, we find a significant contribution to the uncertainties of the fundamental parameters of Kepler-444 at the level of precision of \textit{Kepler} observations. The full results of our study are presented in \cite{Buldgen2019}. This paper is the first of a series aiming at a detailed characterization of the evolution of the planetary system of Kepler-444, following the methodology of \citet{Privitera2016I,Privitera2016AII,Privitera2016III,Meynet2017,Rao2018}.

In addition, we show the need for non-standard ingredients to fully reproduce seismic observations by \textit{Kepler} for this well-known planet host star. 


\section{The solar modelling problem and its contributors}

The solar modelling problem, resulting from the revision of the solar abundances by \citet{AGSS09}, has been the subject of long standing debates in the stellar modelling community (see e.g. \citet{Bahcall2005a, Montalban06, Antia05, Guzik2008,Basu08} and references therein). To this day, no univoqual solution to the issue has been found, as multiple contributors can be at play to explain the observed disagreement between the new generation of standard solar models and helioseismic constraints.

\begin{figure}[ht!]
 \centering
 \includegraphics[width=0.7\textwidth,clip]{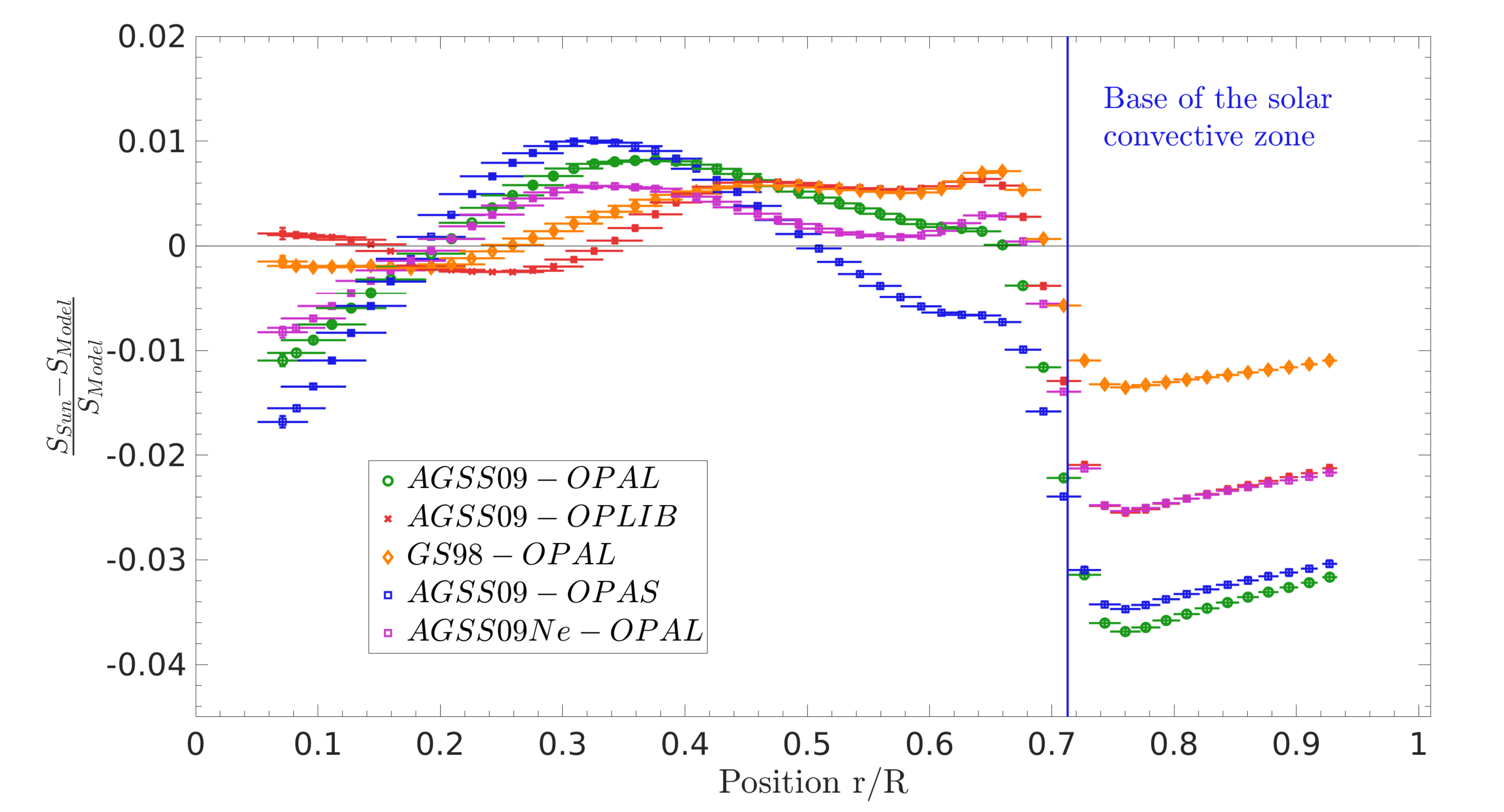}      
  \caption{Inversion of the entropy proxy for various standard solar models, built varying the opacity and chemical abundance tables.}
  \label{buldgen:fig1}
\end{figure}

In Fig. \ref{buldgen:fig1}, we illustrate the impact of varying the opacity tables and abundance tables on the inversion of the entropy proxy defined in \citet{BuldgenS}. 

These modifications are only few of many examples of physical ingredients that are uncertain. Others include: the chemical mixing at the base of the convective envelope, strongly related to the missing dynamical processes in standard solar models; recently, \citet{Zhang2019} put forward the combined effects of such additional mixing to early accreation and mass-loss to erase the discrepancies; \citet{Ayukov2017} computed modified solar models in extended calibration procedures by allowing opacity modifications and nuclear reaction rates modification, leading to similar results. \citet{BuldgenSun} carried out an extensive comparison varying the equation of state, opacity tables, formalism for microscopic diffusion and convection and concluded on the necessity for a combination of opacity increase and mixing at the base of the convective zone. Similar results were obtained by \citet{JCD18} and opacity modifications were foreseen early on as a potential solution. The inadequacy between recently published opacity tables and the experimental measurements of \citet{Bailey, Nagayama2019} seems to also point towards further revision of radiative opacities in the future and sparked intense discussions in the community.

\section{Implication for the modelling of Kepler-444}

Given the very high precision of \textit{Kepler} data, the limitations of stellar seismic modelling for the best targets are not only related to observational uncertainties but rather to the shortcomings of stellar models. In the case of Kepler-444, various existing studies have been carried out to determine its fundamental properties. In this section, we compute a full seismic modelling of Kepler-444 and discuss the implication of modifying the ingredients of the models on the final precision of the modelling results. We combine seismic data from \citet{Campante2015} to revised spectroscopic parameters \citep{Mack2018} and GAIA DR2 parallaxes in our revised modelling \citep{Evans2018, GAIA2018}.

\subsection{Tests of standard physics}

The details of the modelling procedure are given in \citet{Buldgen2019}, we recall here only a few key points. The first step of the modelling is carried out assuming a given set of physical ingredients and seismic and non-seismic constraints, using the AIMS software \citep{Rendle2019} to compute the relevant probability distributions for the model parameters. 

The second step of the seismic modelling procedure is carried out using a Levenberg-Marquardt minimization technique to test the variations in the optimal stellar parameters under the effects of changes of the model physical ingredients. 

\begin{figure}[ht!]
 \centering
 \includegraphics[width=0.5\textwidth,clip]{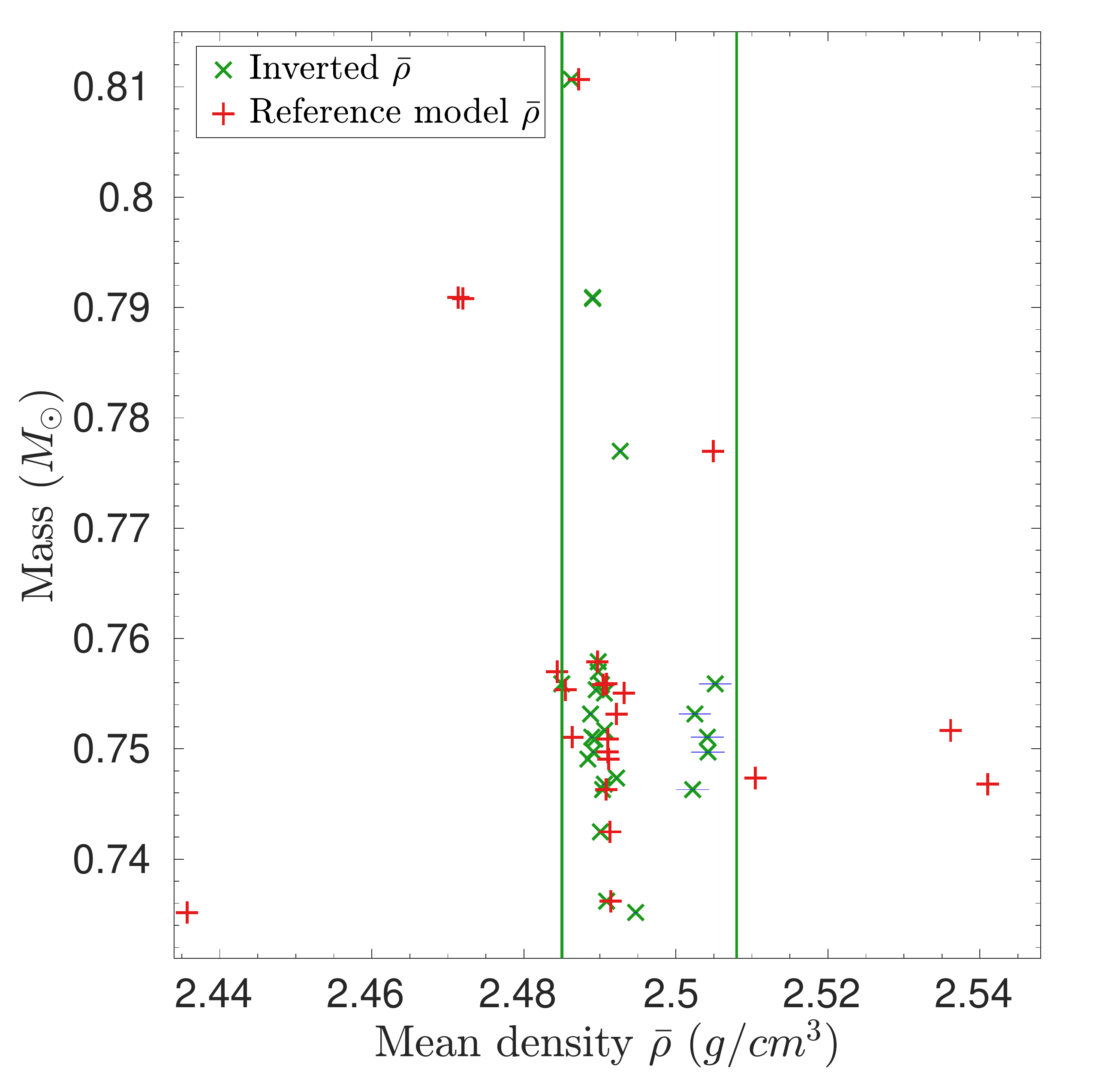}      
  \caption{Results of mean density inversions for the optimal Kepler-444 models in our study.}
  \label{buldgen:fig2}
\end{figure}

As a third step, we carry out seismic inversions of the mean density to further constrain the mass range of acceptable models of our sample. The results of the seismic inversion procedure are illustrated in Fig. \ref{buldgen:fig2}. It is  clear from this plot that the main dispersion of the results stems from the impact of the reference model and of the surface effect correction (see \cite{Buldgen2019} for a detailed discussion). This is a general result of linear seismic inversions based on individual frequencies as shown in \citep{Buldgen2015, BuldgenCygA, BuldgenCygB, Buldgen2017, Buldgen2018}).

Based on this inversion procedure, we have built a reduced sample of reference models and obtained the following values for the mass, radius and age estimates of Kepler-444: $0.754 \pm 0.030$ $\rm{M}_{\odot}$, $0.753 \pm 0.010$ $\rm{R}_{\odot}$ and $11.00 \pm 0.8$ Gy. These uncertainties are slightly lower than those of \cite{Campante2015} who used multiple pipelines to determine the stellar fundamental parameters. Our study thus demonstrates the need to take such biases into account when carrying out detailed modelling of \textit{Kepler} targets. Such aspects are also of paramount importance if seismic inversions of the structure are undertaken, as the trade-off between precision and accuracy is a key factor to avoid overinterpretations of the results. 

\subsection{Non-standard processes acting in Kepler-444}

In addition to testing ``standard'' ingredients of stellar models, we also observed that a significant improvement of the agreement with the frequency ratios could be obtained when taking into account convective overshooting during the evolution of the transitory convective core at the beginning of the main-sequence. The inclusion of overshooting allows for $^{3}\rm{He}$ out-of-equilibrium burning for up to $8$ Gy. As the temperature dependency of out-of-equilibrium burning of $^{3}\rm{He}$ is much higher than that of the equilibrium burning, a convective core is maintained throughout this phase. This was already seen in the CoRoT target HD203608 which had kept a convective core up to its present age \citep{Deheuvels2010}. 

\begin{figure}[ht!]
 \centering
 \includegraphics[width=0.8\textwidth,clip]{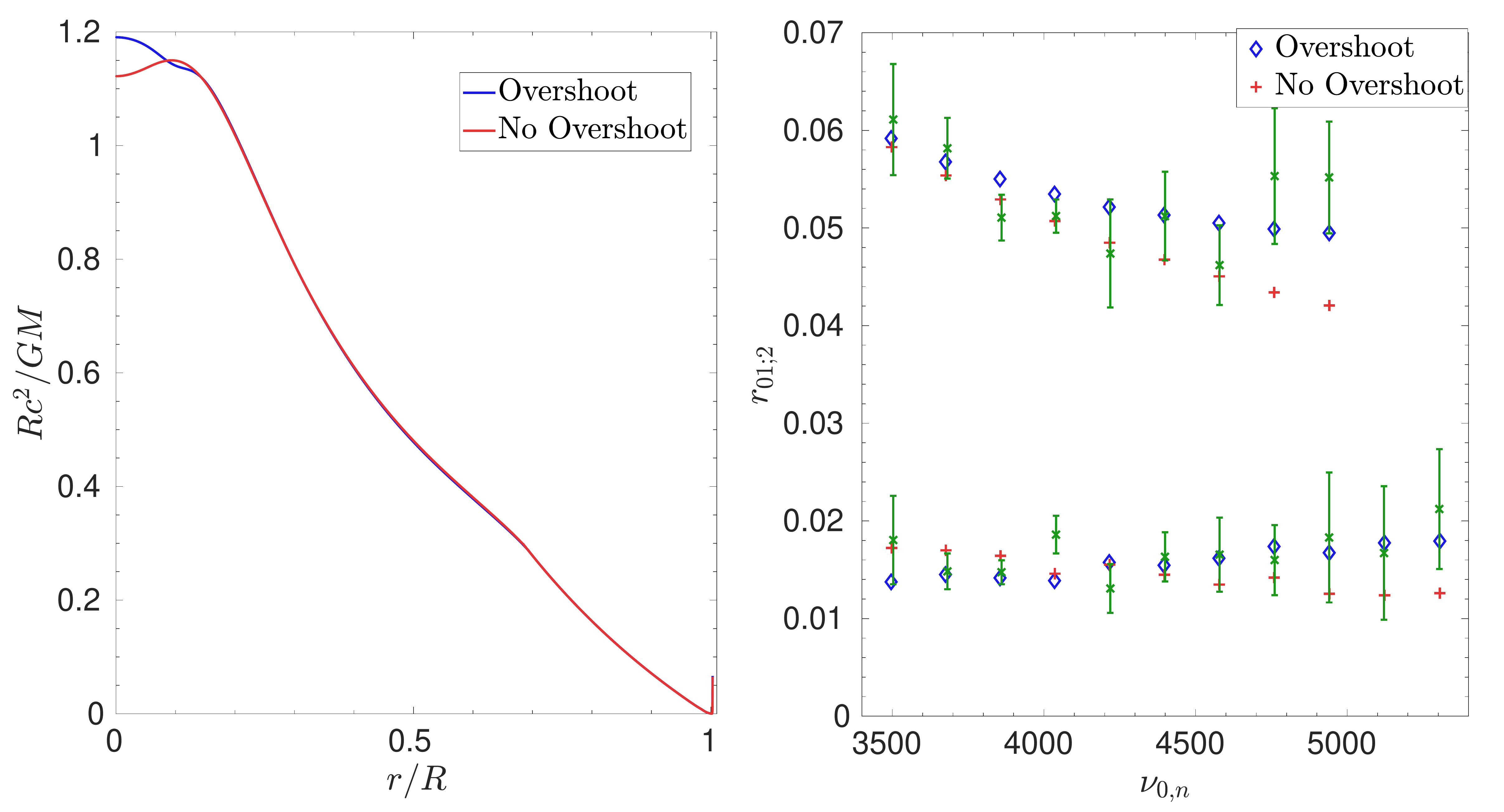}      
  \caption{Left panel: Sound speed profile of models of Kepler-444 built with and without core overshooting. Right panel: Frequency ratios for the models built with and without core overshooting.}
  \label{buldgen:fig3}
\end{figure}

In the case of Kepler-444, the convective core has already disappeared but the sound-speed profile has kept a trace of this phase, as can be seen in the left panel of Fig. \ref{buldgen:fig3}. This is then seen in the frequency ratios (as defined in \cite{RoxburghRatios}) of Kepler-444, which are better reproduced when the overshooting is included, as shown in Fig. \ref{buldgen:fig3}. This result is independent of the ``standard'' input physics used for the model. 

\begin{figure}[ht!]
 \centering
 \includegraphics[width=0.7\textwidth,clip]{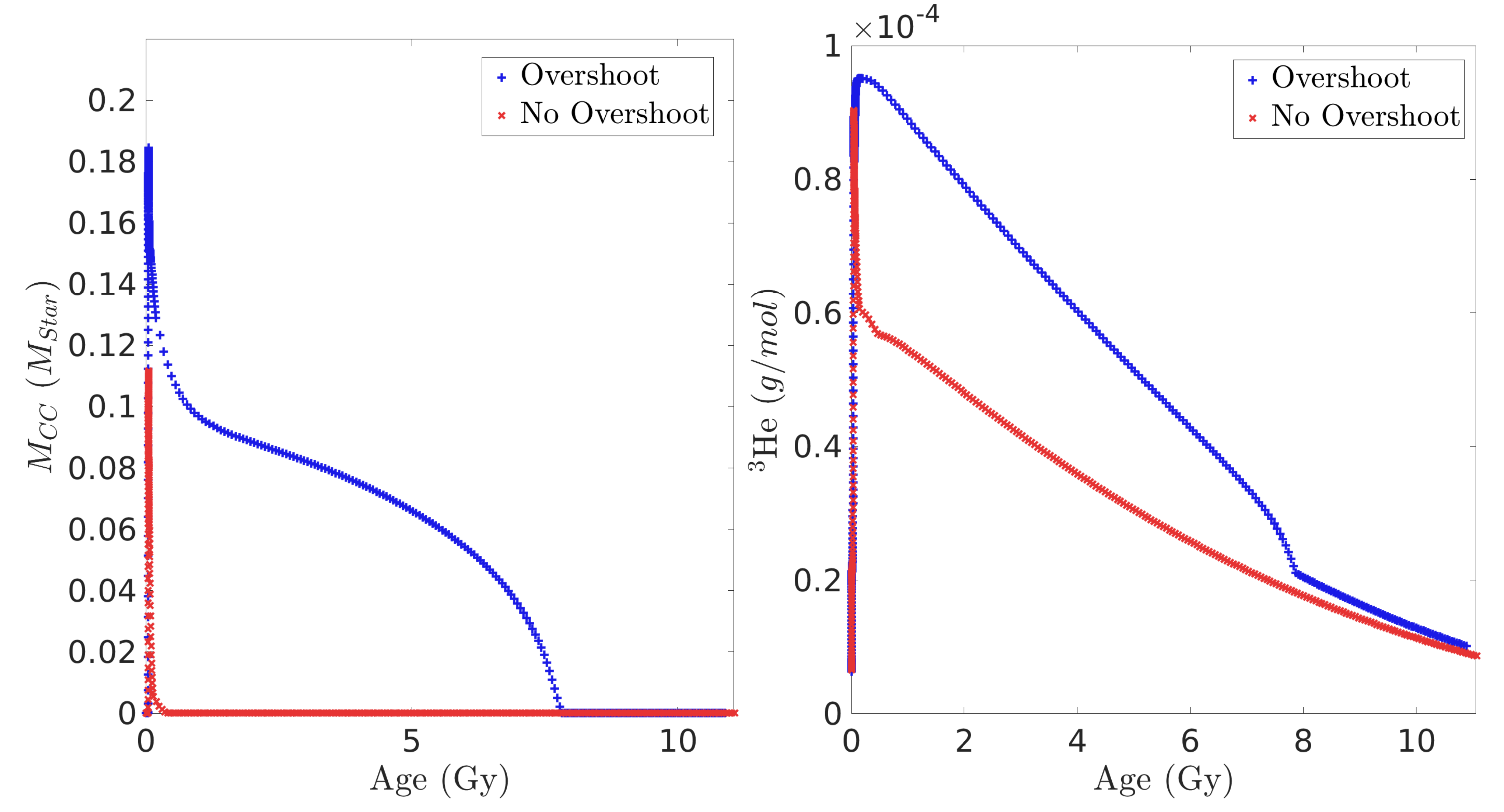}      
  \caption{Left panel: evolution of the mass of the convective core as a function of age for models of Kepler-444 with and without overshooting. Right panel: Central abundance of $^{3}\mathrm{He}$ for models of Kepler-444 with and without overshooting.}
  \label{buldgen:fig4}
\end{figure}

As seen from Fig. \ref{buldgen:fig4}, the key element is to maintain the convective core by keeping the $^{3}\rm{He}$ abundance above its equilibrium value. Once this condition is not fulfilled, the convective core disappears. Thus, the chemical mixing of $^{3}\rm{He}$ is the key aspect that is missing in the standard models and this may not necessarily be achieved only through convective overshooting, as rotation or even non-linear effects of gravity modes \citep{Sonoi2012} may be at the origin of the mixing. 

\section{Conclusions}

In our study, we have carried out a detailed seismic modelling of Kepler-444, building on our knowledge of the solar modelling problem to characterize the spread in fundamental parameters occurring from the uncertainties on key physical ingredients. 

We used a combination of global and local minimization techniques supplemented by inversions of the stellar mean density to determine precise and accurate stellar fundamental parameters for Kepler-444. 
In addition, we have shown that this low-mass star bore a convective core during a significant fraction of its life, namely 8 out of 11 Gy. 

We have shown that the presence of the convective core is a result of out-of-equilibrium burning of $^{3}\rm{He}$, maintained through the additional mixing, bringing fresh fuel in the deep layers. While the presence of mixing seems required to fit the seismic data, its nature is still to be determined. In that sense, the very high quality of seismic data provided by space-based photometry missions provides an unprecedented opportunity to test the limitations of standard stellar models. In parallel, our study shows that not taking such limitations and the resulting biases into account may lead to clear overestimations of the precision of seismic modelling. 

\begin{acknowledgements}
This work is sponsored by the Swiss National Science Foundation (project number $200020-172505$). M.F. is supported by the FNRS (``Fonds National de la Recherche Scientifique'') through a FRIA (``Fonds pour la Formation à la Recherche dans l'Industrie et l'Agriculture'') doctoral fellowship.  S.J.A.J.S. is funded by the Wallonia-Brussels Federation ARC grant for Concerted Research Actions. This article used an adapted version of InversionKit, a software developed within the HELAS and SPACEINN networks, funded by the European Commissions's Sixth and Seventh Framework Programmes.
\end{acknowledgements}

\bibliographystyle{aa}  
\bibliography{buldgen_5o02} 

\end{document}